\documentclass[aps,pre,twocolumn,showpacs]{revtex4}%endfloats
\usepackage{graphicx}%
\usepackage{dcolumn}
\usepackage{amsmath}   
\usepackage[usenames,dvipsnames]{pstricks}
\usepackage{pst-grad} % For gradients
\usepackage{pst-plot} % For axes
\usepackage{pifont,color}
\usepackage{bm}
\usepackage{longtable}
\usepackage{psfrag}

\begin{document}

\title{Dynamical and orientational structural crossovers in low-temperature glycerol} 
\author{Salman Seyedi}
\author{Daniel R.\ Martin}
\author{Dmitry V.\ Matyushov}
\email{dmitrym@asu.edu}
\affiliation{Department of Physics and School of Molecular Sciences, 
         Arizona State University, PO Box 871504, Tempe, Arizona 85287}
\begin{abstract}
Mean square displacements of hydrogen atoms in glass-forming materials and proteins, as reported by incoherent elastic neutron scattering, show kinks in their temperature dependence. This crossover, known as the dynamical transition, connects two approximately linear regimes. It is often assigned to the dynamical freezing of subsets of molecular modes at the point of equality between their corresponding relaxation times and the instrumental observation window. The origin of the dynamical transition in glass-forming glycerol is studied here by extensive molecular dynamics simulations. We find the dynamical transition to occur for both the center of mass translations and the molecular rotations at the same temperature, insensitive to changes of the observation window. Both the translational and rotational dynamics of glycerol show a dynamic crossover from the structural to a secondary relaxation at the temperature of the dynamical transition. A significant and discontinuous increase in the orientational Kirkwood factor and in the dielectric constant is observed in the same range of temperatures. We, however, do not find any indications of a true thermodynamic transition to an ordered low-temperature phase. We therefore suggest that all observed crossovers are dynamic in character. The increase in the dielectric constant is related to the dynamic freezing of dipolar domains on the time-scale of simulations. 
\end{abstract}
\pacs{87.14.E-, 87.15.N-, 87.15.Pc}
\maketitle
\section{Introduction}
\label{sec:1}
Displacements of atoms and molecules induced by thermal agitation generally increase with temperature. A linear growth of the mean-squared displacement (MSD) with increasing temperature is predicted by the Nyquist (fluctuation-dissipation) theorem \cite{Kubo:66,Hansen:03}. The MSD is experimentally extracted from either the intermediate scattering function of the neutron scattering experiment \cite{Gabel:02} or from the fraction of recoilless $\gamma$-ray emission of the $\mathrm{^{57}Fe}$ nucleus in the M{\"o}ssbauer experiment \cite{Parak:03,Young:2011ly}. The Nyquist theorem was found to be violated for a number of glass-forming materials, where a kink in the MSD vs.\ temperature is often observed at the laboratory glass transition \cite{Angell:95}. More complex behavior, with several kinks \cite{Roh:2005ek,Khodadadi:10,Hong:2011qf}, was observed for proteins in partially hydrated powders or in the polycrystalline form \cite{DosterNature:89,Zaccai:00}. 

A typical temperature dependence of the protein MSD starts with the linear increase in accord with the Nyquist theorem and the corresponding vibrational density of states \cite{Zaccai:00,Achterhold:02}. It is followed by one or two low-temperature crossovers and, finally, with a much stronger increase above the temperature of the dynamical transition $T_d\sim 200-250$ K \cite{Doster:08}. This latter temperature depends on a number of factors, including the resolution of the spectrometer, i.e., effectively the time period over which the atomic displacements are recorded \cite{Magazu:2011kz,Fenimore:2013eo}. This phenomenology has attracted significant attention since enhanced flexibility and, therefore, the ability to perform biological function can develop at $T>T_d$ \cite{Schiro:2015cv}. 
 
A somewhat unexpected observation came recently from Capaccioli \textit{et al} \cite{Capaccioli:2012jc}, who presented two key observations based on the analysis of a large database of neutron scattering data accumulated so far: (i) the MSD measured in 50:50 lysozyme-glycerol mixture can be nearly seamlessly overlaid with corresponding measurements for the pure glycerol and (ii) there are two crossover temperatures common to lysozyme-glycerol and glycerol systems, at $T_d\simeq 210$ and 276 K. 

The first observation is significant for assigning the modes of the protein-solvent system responsible for the protein's extended flexibility at high temperatures. High protein flexibility is required for its biological action \cite{Smith:91,Zaccai:03,Parak:03}, and this perspective connects protein function with specific physical modes and fluctuations of the protein-solvent system \cite{DMjpcm:15}. Frauenfelder and co-workers suggested that the solvent mode coupled to the protein atomic displacements has to be attributed to the hydration shell \cite{Fenimore:04,Lubchenko:2005hv}. They also noted that this mode is decoupled from the $\alpha$-relaxation of the bulk solvent (structural or collective relaxation with the longest relaxation time and usually connected to the liquid viscosity). The relaxation time of the hydration shell is both faster than $\alpha$-relaxation and is Arrhenius, with the activation energy usually smaller than that of $\alpha$-relaxation.  Taken together, these features point to its $\beta$-character in the established classification of glass science \cite{AngellJAP:00,Donth:01}. Since secondary $\beta$-relaxation processes exist also in the bulk solvent, the fluctuations localized in the hydration shell of the protein are classified as $\beta_h$-relaxation and are expected to carry the dynamics distinct from the bulk \cite{ChenPM:08}. The dynamical transition then occurs when the $\beta_h$-relaxation of the hydration shell slows sufficiently down, with lowering temperature, to become longer than the instrumental time-scale (dynamical freezing) \cite{Khodadadi:08,Frauenfelder:09}.  

The observation of a near-equivalence of MSDs recorded by neutron scattering in lysozyme-glycerol and pure glycerol systems puts under question the hydration-shell hypothesis, or at least the part of it attributing $\beta$-relaxation specifically to the shell, in contrast to a faster relaxation mode of the bulk (of presumably $\beta$-character). The question posed by this observation is whether the modes of the solvent coupled with protein flexibility are hydration-shell specific or generic to the bulk material. Furthermore, since the dynamical transition is a general phenomenon common to glass-forming materials, including molecular liquids and biopolymers \cite{Angell:95}, the question here is what are the modes that experience dynamical freezing at $T_d$ and whether the instrumental resolution must necessarily be a part of the explanation. Addressing some of these mechanistic questions is a goal of this study.

In order to avoid the complexities of protein solutions, we address these basic questions by focusing solely on bulk glycerol, for which we report here extensive molecular dynamics (MD) simulations. The temperature dependence of hydrogen MSDs is analyzed in terms of separate contributions of the center of mass translations and rotations relative to the center of mass of the molecule. Both translational and rotational MSDs show a crossover at the same temperature $T_d\sim 275$ K consistent with experimental data. The temperature of translational and rotational dynamical transitions does not change when the observation time is significantly altered. We also find that the same temperature characterizes the dynamic crossover from $\alpha$ to $\beta$ relaxation as measured by glycerol's diffusivity and rotational dynamics. 

The consistent picture arising from our observations is that a structural crossover occurs in glycerol at $\sim 250-275$ K, which affects both the MSDs and relaxation times. However, there is no indication from our data that this crossover should be identified with a true thermodynamic transition. We therefore suggest that all observed crossovers are dynamical in character. In particular, the structural crossover to a low-temperature state of glycerol, characterized by long-ranged dipolar correlations, becomes possible because these collective correlations cannot relax on the limited observation time. The dynamical transition in the MSD recorded by neutron scattering is not the result of crossing of the time-scale of single-particle translational/rotational diffusion with the observation time-scale, but rather the crossing of the latter with the time-scale of multi-body relaxation of polarized  domains. A corresponding significant increase in the orientational Kirkwood factor and the jump in the dielectric constant at low temperatures are caused, in our simulations, by the crossing of the relaxation time of dipolar domains and the observation (simulation) time. This phenomenology is similar to that of relaxor ferroelectrics where dynamic freezing of ferroelectric domains is responsible for the high dielectric constant of the low-temperature phase \cite{Samara:03}.

\section{Incoherent neutron scattering}
\label{sec:2} 
The experimental MSDs are extracted from incoherent elastic neutron scattering. The reported signals are affected by the instrumental resolution function convoluting with the dynamic structure factor $S(q,\omega)$, for which we assume the scattering momentum $\mathbf{q}$ directed along the $x$-axis of the laboratory frame. The function $S(q,\omega)$ is the time Fourier transform of the intermediate scattering function 
\begin{equation}
I(q,t) = N^{-1}\sum_j \langle e^{iq\Delta x_j(t)}\rangle ,
\label{eq:1}
\end{equation}
where $\Delta x_j=x_j(t)-x_j(0)$ is the displacement of a hydrogen atom and the sum runs over $N$ hydrogen atoms in the system; $\langle \dots\rangle$ denotes an ensemble average. 

In what follows we will consider all hydrogens in the system identical, although we will separate two groups of hydrogens of glycerol: 3 hydroxyl hydrogens and 5 hydrogens bonded to carbon atoms. Correspondingly, experimental results for partially deuterated glycerol \cite{Wuttke:1995vm} $\mathrm{C_3H_5(OD)_3}$ (g-d3) and $\mathrm{C_3D_5(OH)_3}$ (g-d5) will be analyzed by considering the corresponding groups of hydrogen atoms not substituted by deuteration.  

The intensity of the elastic scattering function at $\omega=0$ gives access to the MSD \cite{Gabel:02,Wuttke:1995vm}. The corresponding function $S(q,\omega=0,\Delta\omega)$, depending on the resolution window of the spectrometer $\Delta \omega$, can be approximated by $I(q,t_r)\simeq S(q,\omega=0,\Delta\omega)$, where the resolution time $t_r$ is related to the resolution window of the spectrometer. According to Doster \textit{et al} \cite{Doster:2003kn}, the connection is $t_r/\mathrm{ps} =1.09/\Gamma \mathrm{(meV)}$, where $\Gamma$ is the width at half maximum of the resolution function.   

The intermediate scattering function in Eq.\ \eqref{eq:1} can be estimated 
in the Gaussian approximation \cite{Yi:2012du}, which leads to 
\begin{equation}
-\ln\left[I(q,t)\right] \simeq  q^2\langle (\delta x)^2 \rangle - q^2 \langle \delta x(t)\delta x(0)\rangle  .
\label{eq:2}
\end{equation}
If the time autocorrelation function $\langle \delta x(t)\delta x(0)\rangle$, $\delta x(t)=x(t) - \langle x\rangle$  decays sufficiently to zero on the resolution time $t_r$, the second term in Eq.\ \eqref{eq:2} disappears and one gets an estimate of the mean square fluctuation (MSF) $\langle (\delta x)^2\rangle$ from the linear slope of $-\ln(I(q,t_r))$ vs $q^2$ \cite{Wuttke:1995vm,Schiro:2012fk}. Otherwise one obtains half of the MSD $(1/2)\langle \Delta x(t_r)^2\rangle$ from the slope of $-\ln(I(q,t_r))$ vs $q^2$.

\begin{figure}
\includegraphics*[clip=true,trim= 0cm 1.5cm 0cm 0cm,width=7cm]{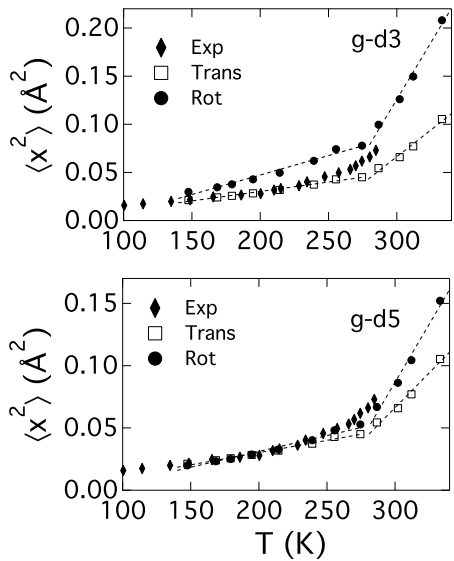}
\caption{$\langle x^2\rangle=\langle \Delta x(t_r)^2\rangle$ for g-d5 (upper panel) and g-d3 (lower panel). The experimental data obtained from IN13 spectrometer for correspondingly deuterated glycerol \cite{Capaccioli:2012jc} are compared to MD simulations. The simulated MSDs are separated into displacement of the glycerol center of mass (``Trans'') and the displacements of hydrogens relative to the center of mass  (``Rot''). The dashed lines are the linear regressions drawn through the corresponding points from MD simulations.  
}
\label{fig:1}
\end{figure}

The data presented here were obtained from extensive MD simulations of glycerol described by the OPLS-AA force field \cite{Caleman:2012fp} as explained in Appendix \ref{secA1}. Our main purpose in the analysis of the intermediate scattering function is to extract the relative contributions to the observed MSD arising from center of mass translations and molecular rotations relative to the center of mass. The question that we address here is whether the dynamical transition, if observed, occurs at the same temperature for these two modes. In addition to general mechanistic insights that such an analysis can produce, this question is relevant to testing the idea of dynamical freezing of a subset of molecular motions as the reason for the experimentally observed kink in the dependence of the MSD on temperature \cite{DosterNature:89,Doster:08,Zaccai:00,Magazu:2011kz}, identified with $T_d$. If the kink is caused by reaching the equality between the relaxation time and the instrumental observation window \cite{Magazu:2011kz}, the dynamical transition temperature should be different for translations and rotations having their distinct relaxation times, unless they happen to be close. This is not what we observe from our simulations: the dynamical transition temperatures are the same for rotations and translations when calculated from fitting the intermediate scattering function to Eq.\ \eqref{eq:2} (Fig.\ \ref{fig:1}). 

The separation of the center of mass translations and rotations relative to the center of mass assumes the factorization of the intermediate scattering function into the translational, $I_T(q,t)$, and rotational, $I_R(q,t)$, components 
\begin{equation}
I(q,t) = I_T(q,t) I_R(q,t) .
\label{eq:5}
\end{equation}
We therefore calculated $I_T(q,t)$ and $I_R(q,t)$ separately and produced the linear fits of the corresponding functions vs $q^2$ with $t_r=25$ ps for both g-d3 and g-d5 liquids. No deuteration was actually performed in simulations and only the corresponding groups of hydrogen atoms were selected to produce the intermediate scattering functions.  

\begin{figure}
\includegraphics*[clip=true,trim= 0cm 1.2cm 0cm 0cm,width=7cm]{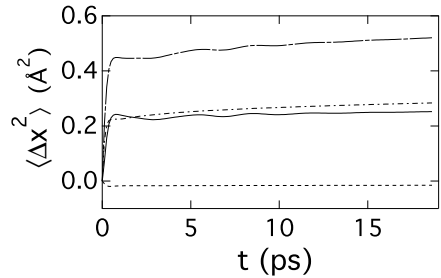}
\caption{$\langle \Delta x(t)^2\rangle$ vs time at $T=250$ K. The overall MSD (long-dashed line) is separated into the center of mass (solid line), rotational (dash-dotted line), and cross (dashed line) components (Eq.\ \eqref{eq:6}).     
}
\label{fig:2}
\end{figure}

The accuracy of translation/rotation factorization in Eq.\ \eqref{eq:5} was tested previouly and is usually found to hold \cite{Teixeira:1985vk,Chen:1997ts,Liu:2002kx}. Indeed, one expects this separation to be accurate in the Gaussian limit since translations and rotations carry different symmetry. If one separates $\Delta x(t)=\Delta x_c(t) + \Delta x_R(t)$ into the center of mass displacement $\Delta x_c(t)$ and the rotation relative to the center of mass $\Delta x_R(t)$, the MSD becomes the sum of two self terms and the translational-rotational cross term 
\begin{equation}
  \langle\Delta x(t)^2\rangle = \langle \Delta x_c(t)^2\rangle + \langle \Delta x_R(t)^2\rangle + 2\langle \Delta x_c(t)\Delta x_R(t)\rangle .
  \label{eq:6}
\end{equation}
Figure \ref{fig:2} shows an example of the analysis of the three correlation components in Eq.\ \eqref{eq:6} from MD simulations. The translational and rotational components of the MSD are close in magnitude, while the cross-correlation is negative and is much smaller. 

The translational and rotational MSDs are shown separately in Fig.\ \ref{fig:1} to indicate the common point of the kink at $T_d\sim 275$ K. The same temperature of the dynamical transition is reported experimentally \cite{Wuttke:1995vm,Capaccioli:2012jc}. However, the absolute values of MSDs from experiment (closed diamonds in Fig.\ \ref{fig:1}) are below the simulation results, which is easy to see from the plot since the overall MSD follows from adding up the translational and rotational components (Eq.\ \eqref{eq:6}). The most probable explanation of this discrepancy is that fitting the experimental neutron scattering data in a limited range $q$-values used in the measurements \cite{Wuttke:1995vm} allows one to probe only a limited subset of motions \cite{Chen:1999uj,Liu:05}, presumably the translational diffusion. Indeed, the agreement between simulations and experiment for the center of mass MSD is quite good. We also note that the agreement between the calculated coefficient of self-diffusion of glycerol and the results of measurements by NMR \cite{Chen:2006jv} is also reasonable (Fig.\ \ref{fig:6} below).

\begin{figure}
\includegraphics[clip=true,trim= 0cm 1.2cm 0cm 0cm,width=7cm]{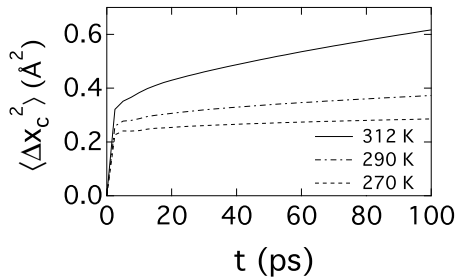}
\caption{Center of mass MSD, $\langle \Delta x_c(t)^2\rangle$, of glycerol at different temperatures indicated in the plot.}
\label{fig:3}
\end{figure}

\begin{figure}
\includegraphics*[clip=true,trim= 0cm 1.5cm 0cm 0cm,width=7cm]{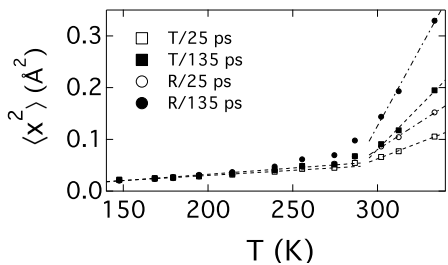}
\caption{$\langle x^2\rangle=\langle \Delta x(t_r)^2\rangle$ for g-d5 measured on $t_r=25$ ps (open points) and $t_r=135$ ps (closed points). The center of mass $\langle\Delta x_c^2$ (``T'', squares) and rotational $\langle\Delta x_R^2$ ("R", circles) contributions are shown separately. The dashed and dash-dotted lines are linear regressions drawn through the low-temperature and high-temperature points.         
}
\label{fig:4}
\end{figure}

The time dependence of MSDs shown in Fig.\ \ref{fig:2} also helps to understand the physical origin of MSDs recorded by neutron scattering. Both the translational and rotational components of the MSDs are characterized by two distinct regimes: a fast  ($\sim 1$ ps)  growth due to ballistic motions in the liquid's cage (localized diffusion \cite{Mamontov:2012tg}), followed by a much slower, long-range diffusion with $\langle \Delta x(t)^2\rangle \propto t$. The main observation here is that most of the MSD on the resolution time-scale $t_r\sim 25$ ps is caused by the ballistic displacement associated with a secondary relaxation and not by the diffusional motion associated with the primary relaxation process. This conclusion holds both below and above $T_d$ (Figs.\ \ref{fig:2} and \ref{fig:3}). The increase of the observation window from 25 ps to 135 ps makes the time spent by the particle on the linear, diffusion portion of the MSD longer (Fig.\ \ref{fig:2}) and thus increases the slope of the high temperature part of the MSD curve (Fig.\ \ref{fig:4}). It is important to realize that fast cage dynamics, resulting to the main portion of the observed MSD, are much faster than the resolution time $t_r$ and in fact become even faster with lowering temperature because of a greater rigidity of the low-temperature glycerol. It is the amplitude of the ballistic displacement which gets larger with increasing temperature, resulting in the observed temperature dependence of the MSD. The crossing of the resolution time of the spectrometer (25 ps) and the relaxation time of these ballistic motions never occurs and, therefore, the kink in the MSD vs temperature cannot be attributed to the finite resolution time.

The change of the form of the MSD vs $T$ with the changing observation window $t_r$ is shown in Fig.\ \ref{fig:4}. It adds additional evidence to the suggestion that the kink in the MSD's temperature dependence is not caused by the equality between the relaxation time and the observation window. While the high-temperature portion of the MSD has a steeper slope for a higher $t_r$, in agreement with experiment \cite{Capaccioli:2012jc}, the temperature of the dynamical transition $T_d$ has little sensitivity to $t_r$. In addition, the equality between the dynamical transition temperatures for the translational and rotational MSDs is preserved between $t_r=25$ ps and $t_r=135$ ps. If one assumes that the consistency in $T_d$ for $t_r=25$ ps shown in Fig.\ \ref{fig:1} is a mere coincidence, it is hard to see how it can be preserved at $t_r=135$ ps. One has to accept the conclusion that the kink in the MSD is not related to the observation window \cite{Schiro:2012fk,Fomina:2014cp} and, instead, should be attributed to the softening of the liquid cage, with increasing temperature, in which a glycerol molecule finds itself for a relatively short time of $\sim 1$ ps. The rattling inside the cage is followed by an escape and the onset of long-range diffusion, but this component simply adds to the main displacement achieved by the ballistic cage rattling. The next question is whether structural distinctions of the entire liquid producing the difference between the low-temperature rigid cage and the high-temperature soft cage can be identified.

\section{Dynamic crossover}
\label{sec:3} 
An explanation alternative to the instrumental resolution effect for the appearance of the kink in the proton MSD involves the dynamic crossover, i.e., a corresponding kink in the dependence of the system relaxation time on the inverse temperature \cite{ChenPNAS:06}. This phenomenology, known as the fragile-to-strong transition in glass science \cite{AngellJAP:00}, represents the crossover from the structural $\alpha$-relaxation at high temperatures above the crossover to a secondary $\beta$-relaxation at low temperature below the crossover. Correspondingly, the activation barrier of the high-temperature $\alpha$-relaxation is higher than the activation barrier of the low-temperature $\beta$-relaxation. We show below that this phenomenon is not connected to the kink in the MSD reported by neutron scattering and, at least in our simulations, has a trivial explanation of slower dynamics exceeding in its relaxation time the observation window (simulation time in the case of MD).  

The problem of dynamic crossover in confined water has been extensively studied \cite{Zanotti:05,Liu:05,Cupane:2014hk} and it has been established that the  temperature of the dynamic crossover of confined water is generally consistent with $T_d$ of proteins \cite{ChenPNAS:06,Schiro:2013gc}. The temperature $T_d$ was also found to be independent of the protein hydration level  \cite{Schiro:2013gc,Fomina:2014cp,Mallamace:2015ea} even though the relaxation times themselves are strongly affected by hydration. This latter observation points to the connection between $T_d$ and some sort of structural change in confined water. 

The dynamic crossover results for water are necessarily limited to confined systems since bulk water is unstable to nucleation below $\simeq 243$ K \cite{Mallamace:06,Cupane:2014hk}. Since our present simulations apply to bulk glycerol, it would be of significant interest to establish a phenomenology similar to that found for confined water for a material available in bulk phase both in simulations and in the laboratory experiment.     

\begin{figure}
\includegraphics*[clip=true,trim= 0cm 1cm 0cm 0cm,width=7cm]{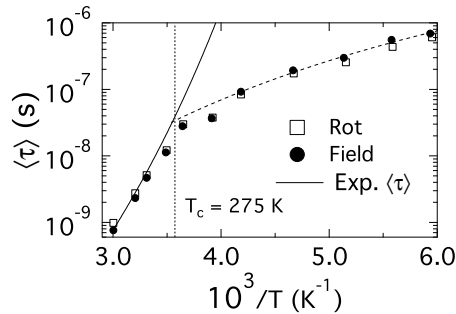}
\caption{Average relaxation time $\langle \tau\rangle$ obtained from rotational correlation function (``rot'') and from the electric field correlation function (``field''). The solid line refers to the average relaxation time \cite{Richert:2014wa} obtained by fitting the dielectric loss spectrum to the Cole-Davidson function \cite{Menon:92}. The dashed line is a regression drawn through the MD points obtained from the electric field correlation function. $T_c$ (dotted line) indicates the crossover temperature.   
}
\label{fig:5}
\end{figure}

It is useful to start off with an estimate of how the dynamic crossover in the relaxation time can potentially affect the MSD measured on the resolution time $t_r$. This can be illustrated for the rotational MSD, which can be rewritten in terms of the rotational MSF $\langle (\delta x_R)^2\rangle=\langle x_R^2\rangle - \langle x_R\rangle^2$ and the normalized autocorrelation function of rotations $\phi_R(t)$ 
\begin{equation}
\langle\Delta x_R(t)^2\rangle =  2\langle(\delta x_R)^2 \rangle\left[1- \phi_R(t) \right], 
\label{eq:7}
\end{equation}
where 
\begin{equation}
\phi_R(t) = \langle(\delta x_R)^2\rangle^{-1} \langle\delta x_R(t)\delta x_R(0)\rangle .   
\label{eq:8}
\end{equation}
The generic form of $\phi_R(t)$ is the initial ballistic (Gaussian) decay, followed by exponential collective relaxation: $\phi_R(t)=A_g\exp[-(t/\tau_g)^2] + (1-A_g)\exp[-t/\tau_R]$  (or, alternatively, multi-exponential or stretched exponential term) \cite{Hansen:03}. In the entire temperature range studied for glycerol we find that $t_r$ falls between the time of ballistic relaxation $\tau_g$ and the time of collective exponential relaxation $\tau_R$: $\tau_g\ll t_r\ll \tau_R$. One therefore gets
\begin{equation}
\langle\Delta x_R(t)^2\rangle \simeq  2\langle(\delta x_R)^2 \rangle \left[A_g+(1-A_g)(t_r/\tau_R) \right].
\label{eq:9}  
\end{equation}
In the limit of $t_r \ll \tau_R$, the relaxation time is not expected to affect the MSD. Its magnitude is mostly determined by the amplitude of the Gaussian component of the relaxation dynamics, in agreement with the arguments presented in relation to Figs.\ \ref{fig:2} and \ref{fig:3}. Therefore, if the dynamic crossover and the kink of the MSD occur at the same temperature \cite{Schiro:2013gc} one has to relate this coincidence to a structural change and not to a direct effect of the relaxation time on the MSD. The hypothesis that the crossover in the relaxation time affects the MSD is, therefore, not supported by our simulation results.  

\begin{figure}
\includegraphics*[clip=true,trim= 0cm 1cm 0cm 0cm,width=7cm]{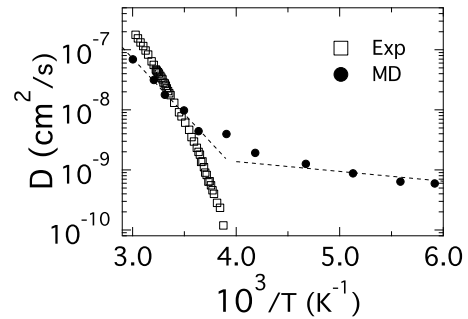}
\caption{Diffusion coefficients recorded experimentally by NMR (``Exp'', \cite{Chen:2006jv}) and obtained from the simulations (``MD''). The dashed line is a regression drawn through the MD points.    
}
\label{fig:6}
\end{figure}

The results for the average rotational relaxation time for all protons in glycerol are shown in Fig.\ \ref{fig:5}. It is calculated by integrating the time correlation function  
\begin{equation}
\langle \tau_X \rangle =  \int_0^\infty \phi_X(t) dt,
\label{eq:10}
\end{equation}
where $X=R$ corresponds to the normalized time correlation function in Eq.\ \eqref{eq:8}. These results are shown by the open points in Fig.\ \ref{fig:5}.

We have additionally calculated the time correlation function $\phi_E(t)\propto\langle \delta\mathbf{E}(t)\cdot \delta\mathbf{E}(0)\rangle$ based on the dynamic variable of the electric field produced by the rest of the glycerol liquid at the center of mass of a given target molecule ($\mathbf{X}=\mathbf{E}$).  The microscopic electric field $\mathbf{E}(t)$ is therefore a fluctuating local field producing a torque on the glycerol's dipole moment. The results for the average relaxation times obtained from the corresponding time correlation functions through Eq.\ \eqref{eq:10} are shown by the closed points in Fig.\ \ref{fig:5}. There is a good agreement between $\tau_R$ and $\tau_E$ suggesting that the electric field fluctuations are caused by molecular rotations, as one would anticipate from the standard Debye model of dielectric relaxation \cite{Frohlich,Richert:2014wa}. 

The average relaxation times from MD simulations  are compared in Fig.\ \ref{fig:5} with the average relaxation time calculated from the Cole-Davidson fit of glycerol's loss spectrum reported by broad-band dielectric spectroscopy \cite{Menon:92} (solid line). There is a very good agreement between the simulations and experimental dielectric data at high temperatures, suggesting that the adopted force field \cite{Caleman:2012fp}  (see Appendix \ref{secA1}) is well parametrized for glycerol rotations. There is a less satisfactory agreement between the diffusion coefficient calculated from MD and measured by NMR (Fig.\ \ref{fig:6}).  Differences between quasi-elastic neutron scattering (QENS) and NMR/viscosity data for glycerol self-diffusion  have been documented in the past \cite{Cuello:1998wt,Mamontov:2012tg} and might contribute to the discrepancy. 

\begin{figure}
\includegraphics*[clip=true,trim= 0cm 1.5cm 0cm 0cm,width=7cm]{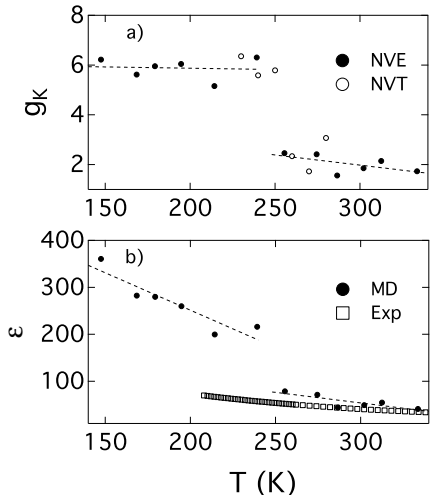}
\caption{The Kirkwood factor (a) and dielectric constant (b) of glycerol calculated from MD (circles) and measured in bulk samples experimentally \cite{DMjcp1:16} (squares). The dashed lines are linear fits to the corresponding subsets of data to guide the eye. The Kirkwood factors in (a) were obtained both in NVE and NVT separate simulation runs. 
}
\label{fig:7}
\end{figure}

The dynamic crossover occurs in the range of temperatures when the $\alpha$-relaxation time becomes comparable to the length of the simulation trajectory $\tau_\text{sim}\simeq 50$ ns. In fact, the time window $\tau_\text{calc}$ on which the time correlation function $\phi_X(t)$ is calculated from the simulation trajectories is always shorter, $\tau_\text{calc}<\tau_\text{sim}$. We therefore stop observing the slow relaxation in simulations when the $\alpha$-relaxation time becomes longer than $\tau_\text{calc}$. The relative weight of the fast relaxation in $\langle \tau\rangle$ increases and we observe this as a dynamical crossover. 

What our data do not seem to address is why the kinks in the rotational and translational MSDs and the corresponding dynamical crossovers in the rotational relaxation times and translational diffusion (Figs.\ \ref{fig:5} and \ref{fig:6}) all occur in the same range of temperatures. A possible scenario to explain this coincidence might include a structural transition resulting in a drop of the configurational entropy \cite{DMjcp1:07}. According to the general arguments based on the Adam-Gibbs relation \cite{AngellJAP:00}, this would result in a much slower main relaxation process, which would sharply disappear from the observation window of our numerical experiment. While our results presented below do support alteration of glycerol's orientational structure, we do not have a direct evidence for a discontinuous change in the configurational entropy.  

In order to identify possible structural changes, we have looked at the temperature dependence of the Kirkwood factor reflecting orientational correlations in the liquid
\begin{equation}
g_\text{K} = \sum_{m} \langle \mathbf{\hat e}_\ell\cdot \mathbf{\hat e}_m\rangle . 
\label{eq:11}
\end{equation}
Here, $\mathbf{\hat e}_m$ are the unit vectors of molecular dipoles (4.6 D in the force field used in our simulations). The Kirkwood factor was in turn used in the Kirkwood-Onsager relation \cite{Frohlich} to calculate the dielectric constant $\epsilon(T)$ (the glycerol force field is non-polarizable and the refractive index is equal to unity). The results of these calculations are shown in Fig.\ \ref{fig:7}.  

\begin{figure}
\includegraphics*[clip=true,trim= 0cm 1.5cm 0cm 0cm,width=7cm]{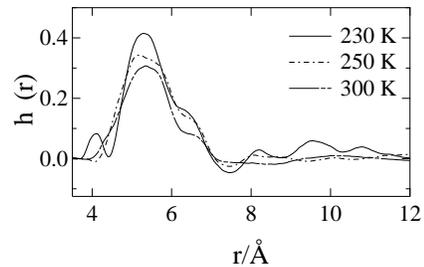}
\caption{{\label{fig:8}}Projection of the pair distribution function of glycerol on the orientational invariant $\Delta(1,2)=(\mathbf{\hat e}_1\cdot\mathbf{\hat e}_2)$ calculated from MD simulations at different temperatures.}
\end{figure}

The Kirkwood factor shows a discontinuous increase at $T<250$ K, which results in the corresponding increase of the dielectric constant calculated from MD simulations. The increase in $g_\text{K}$ is caused by the emergence of long-range orientational correlations of glycerol dipoles at low temperatures, as is illustrated in Fig.\ \ref{fig:8}. We show there the projection of the pair correlation function of glycerol $h(r,\mathbf{\hat e}_1,\mathbf{\hat e}_2)$, depending on the distance $r$ between two molecules and their orientations $\mathbf{\hat e}_1$ and $\mathbf{\hat e}_2$, on the rotational invariant of the scalar product between the unit vectors of the dipole moments $\Delta(1,2)=(\mathbf{\hat e}_1\cdot\mathbf{\hat e}_2)$. The corresponding pair distribution function \cite{Hansen:03} 
\begin{equation}
h^\Delta (r) = \int h(r,\mathbf{\hat e}_1,\mathbf{\hat e}_2) \Delta(1,2) \frac{d\omega_1 d\omega_2}{(8\pi)^2}  
\label{eq:12} 
\end{equation}
at different temperatures in shown in Fig.\ \ref{fig:8}.

\begin{figure}
\includegraphics[clip=true,trim= 0cm 1cm 0cm 0cm,width=8cm]{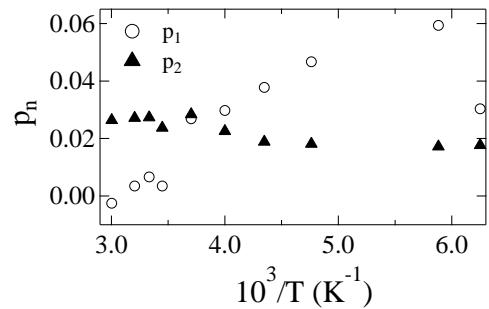}
\caption{Orientational order parameters $p_1$ and $p_2$ in Eq.\ \eqref{eqS9} calculated from MD trajectories at different temperatures. }
\label{fig:9}
\end{figure}

\begin{figure}
\includegraphics[clip=true,trim= 0cm 0cm 0cm 0cm,width=8cm]{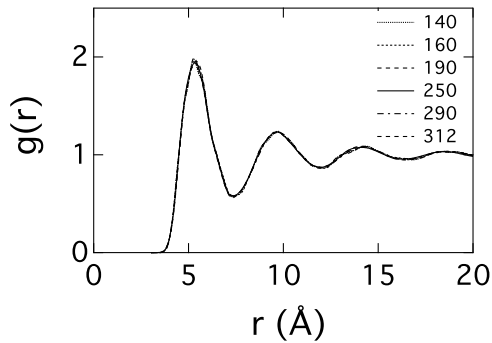}
\caption{Pair distribution functions $g(r)$ of the center of mass of glycerol calculated at the temperatures indicated in the plot. The calculated functions nearly coincide on the scale of the plot. }
\label{fig:10}
\end{figure}

It is clear that a long-range oscillatory pattern, reflecting preferential parallel alignments of the dipoles, appears at low temperatures. The dipolar alignments are responsible for an increase in the low-temperature Kirkwood factor, $g_\text{K}=1 +\rho\int h^\Delta(r) d\mathbf{r}$, $\rho$ is the number density. Despite these long-range orientational correlations, the low-temperature phase does not show any specific orientational order, as confirmed by calculations of the first and second orientational order parameters \cite{Ayton:96,Vertogen:88} (Fig.\ \ref{fig:9}) as explained below.  No translational order is observed either: the radial pair distribution functions are nearly identical at low and high temperatures (Fig.\ \ref{fig:10}). We therefore can conclude that the low-temperature phase is a disordered liquid.      

The orientational order can be detected by orientational order parameters typically defined for liquid crystals \cite{Vertogen:88}. The order parameter $p_n$ is the average $n$th order Legendre polynomial $P_n(\mathbf{\hat e}\cdot \mathbf{\hat n})$ 
\begin{equation}
\label{eqS9}
p_n = N_m^{-1} \sum_\ell \langle P_n(\mathbf{\hat e}_\ell\cdot \mathbf{\hat n}) \rangle 
\end{equation}
relative to the liquid director $\mathbf{\hat{n}}$; $N_m$ is the number of molecules in the liquid. The director is identified as the eigenvector corresponding to the largest eigenvalue of the tensor
\begin{equation}
\label{eqS8}
Q_{\alpha\beta}=(2N_m)^{-1}\sum_\ell \left( 3 \, \hat e_{\ell,\alpha} \hat e_{\ell,\beta} - \delta_{\alpha\beta} \right),
\end{equation}
where $\alpha$ and $\beta$ are the Cartesian projections and $\delta_{\alpha\beta}$ is the Kronecker delta function. The results of calculations for the first and second order parameters ($n=1,2$) are shown in Fig.\ \ref{fig:9}. No orientational order can be identified at low temperatures from these calculations.

The jump in the simulated dielectric constant is in stark disagreement with the linear dielectric experiment \cite{DMjcp1:16} where no discontinuities were observed (squares in Fig.\ \ref{fig:7}b). The results of simulations are in fair agreement with experiment at high temperatures, but the increase in the Kirkwood factor at lower temperatures (Fig.\ \ref{fig:7}a) makes the dielectric constant much higher than observations. Since the crossover temperature for the dielectric constant is roughly consistent with the kinks in the rotational and translational MSDs, we conclude that restricting the observation window not only makes changes to the observable relaxation dynamics, but also does not allow certain orientational correlations to relax. As a result, we observe a long-range orientational order frozen on the observation time-scale. This implies that both the low-temperature Kirkwood factor and the corresponding dielectric constant shown in Fig.\ \ref{fig:7} are non-equilibrium quantities. A similar, about five times compared to the bulk (Fig.\ 9 in Ref.\ \onlinecite{Kasina:2015fp}), increase in the dielectric constant was observed for ultrathin films of glycerol obtained by vapor deposition \cite{Capponi:2012jg}. Subsequent combined dielectric and calorimetry measurements have suggested the existence of rigid polar clusters, which relax as a whole, with an enhanced cluster dipole moment \cite{Kasina:2015fp}. There is also recent evidence of an unrelaxed orientational order in organic glasses obtained by surface deposition \cite{Dalal:2015gt}. 
 
\begin{figure}
\includegraphics*[clip=true,trim= 0cm 0.5cm 0cm 0cm,width=7cm]{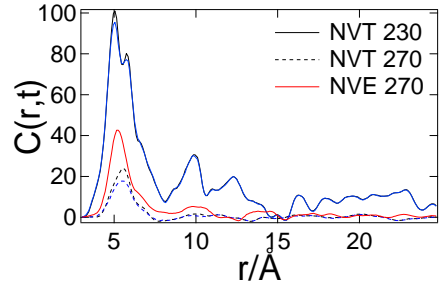}
\caption{$C(r,t)$ at $t=2.5$ ps (black lines) and $t=2.5$ ns (blue lines) calculated from NVT simulations of glycerol at different temperatures indicated in the plot. The red line indicates NVE simulation at 270 K.   
}
\label{fig:11}  
\end{figure}

The existence of highly correlated clusters should be seen in the heterogeneity of binary correlations expressed in terms of fourth-order correlation functions \cite{Berthier:2011hs}. In order to test this hypothesis, we made the next step of calculating the distance- and time-dependent correlations between binary dipolar orientational correlations expressed through the instantaneous Kirkwood factors. Specifically, the quantity
\begin{equation}
c_\ell(t) = \sum_{m\ne\ell} \mathbf{\hat e}_\ell(t)\cdot \mathbf{\hat e}_m(t)   
\label{eq:13}  
\end{equation}
was constructed at each point of the simulation trajectory to reflect the instantaneous binary correlations of the chosen dipole moment $\ell$ with all remaining dipoles in the liquid. Obviously, one has $\langle c_\ell(t)\rangle = g_K - 1$. We then constructed the distance- and time-dependent correlation between the local binary correlations as follows
\begin{equation}
C(r,t) = \frac{V}{N_m^2} \sum_{\ell, k} \langle c_\ell(0) c_k(t) \delta\left(\mathbf{r} - \mathbf{r}_\ell(0) + \mathbf{r}_k(t) \right)\rangle , 
\label{eq:14}   
\end{equation}
where the average is taken along the simulation trajectory and $V$ is the liquid volume. The normalization of $C(r,0)$ relates it to the Kirkwood factor 
\begin{equation}
V^{-1}\int C(r,0) d\mathbf{r} = (5/3) g_K^2 - 2 g_K + 1 . 
\label{eq:15}  
\end{equation}
Similarly to $h^\Delta (r)$ in Fig.\ \ref{fig:8}, but significantly more pronounced, we observe the rise of long-range heterogeneous correlations at low temperatures (Fig.\ \ref{fig:11}).

\section{Discussion and implications for the protein dynamical transition}

We obtained here, by computer simulations, both a kink in the temperature dependence of the MSD (dynamical transition) and the dynamical crossovers in the relaxation times. Both effects have been observed experimentally and a link between them has been suggested through some sort of structural transition in the liquid \cite{ChenPNAS:06,Cupane:2014hk,Mallamace:2015ea}. The answer to the ongoing discussion of whether a purely dynamical crossover or a structural transition explains the data might be that both are present. However, in contrast to the scenarios involving thermodynamic liquid-liquid transitions, both the structural and relaxation time crossovers have a dynamic origin. The structural crossover is caused by the inability of certain structural correlations to relax on the observation window. There is nothing in our data that connects the appearance of such structural correlations to a thermodynamic transition between two phases of a bulk material. This distinction becomes, however, less loaded with physical meaning in the low-temperature state. When the relaxation time of the ``orientationally correlated liquid'' becomes much longer than any conceivable experimental time, one has to distinguish this state of the material as an ``orientationally correlated glass'', with all relevant properties distinct from the ``ordinary'' glass. One arrives at polyamorphism of the glass state \cite{AngellJAP:00} caused by long-ranged orientational correlations.

The observation of an increase in the dielectric constant of glycerol below the dynamical transition, here by simulations and for vapor deposited glasses experimentally \cite{Capponi:2012jg,Kasina:2015fp}, adds a structural component to the standard picture of ergodicity breaking of glass science. The standard paradigm is that the glass does not have the ability to relax, but maintains the structure of the liquid. This is indeed true for the positional structure of the glycerol molecules. However, the inability of dipolar orientations to relax causes orientational heterogeneity represented by correlated dipolar clusters, which do not relax on the observation time-scale. The long-sought growth of the structural order of glass-formers on approach to the laboratory glass transition might be, therefore, best discovered by experiments probing the heterogeneity of orientational multipolar correlations.   

The conclusion that no thermodynamic transformation is at work in creating dipolar domains does not make our observations less ``interesting''. In particular, this scenario is relevant to the role of dynamics and structure of protein's hydration shells in the protein function. About anything related to the protein structure and function has to be described as metastable. Protein itself is unstable to either hydrolysis or association, both bringing it to a thermodynamically more stable state \cite{Scheraga:2015lr}. The function of proteins as enzymes catalyzing specific biochemical reactions is even more affected by the notion of a finite ``observation window'' \cite{DMjpcm:15}. This idea implies that any dynamical or structural information related to the protein itself or to its hydration shell has to be considered from the perspective of a finite observation window provided by the reaction rate, i.e., the characteristic time on which the reactants climb the activation barrier separating them from the products. A dynamic process slower than the rate becomes dynamically frozen and does not contribute to the fluctuation spectrum of the bath driving the reaction. 

The ability of the solvent to preserve a specific structure distinct from its thermodynamic state on a given observation window immediately implies that an enzymetic reaction will ``see'' different solvents, with potentially dramatically different properties (such as polarity), depending on the reaction rate. Figure \ref{fig:7} provides a dramatic confirmation of this possibility showing the ability of glycerol to possess a very high dielectric constant due to its inability to relax its long-range orientational correlations on a given observation window. A related example, with a similar phenomenology, is the appearance of polarized (ferroelectric) domains in the hydration shells of proteins observed on the time-scale of simulations \cite{DMjpcl:15}.  Similarly to our present results for glycerol, these domains might well equilibrate to zero overall dipole on longer time-scales, but a non-zero net dipole of the shell will be recorded by any kinetic process occurring faster than the domain relaxation dynamics. 

Bulk glycerol studied by linear dielectric spectroscopy does not display the features indicative of domain formation. There is a general agreement that linear dielectric spectroscopy does not directly probe heterogeneity of a bulk material \cite{Richert:2014wa}. However, it might still be illuminating to ask why the relaxation of oriented domains in the bulk is not observed by dielectric spectroscopy. One possible answer to this is that the lifetime of a domain is smaller than its rotational relaxation time. The domains dissolve before there is a chance to probe their rotational relaxation. Increasing the lifetime of domains, as potentially achieved by surface vapor deposition \cite{Capponi:2012jg,Kasina:2015fp}, might create conditions for observing the large dipole of the correlated domain.     

The identification of the MSD crossover with the cage dynamics, in the combination with nearly identical behavior of MSD of glycerol and lysozyme-glycerol \cite{Capaccioli:2012jc}, puts under question the need for a special $\beta_h$ relaxation process of the hydration shell \cite{Fenimore:04,Lubchenko:2005hv,Frauenfelder:09} to explain these data. It appears that fast secondary relaxation of bulk glycerol ($\beta_f$ in the standard classification of glass science \cite{Ediger:96,Lunkenheimer:2002fk,Lunkenheimer:2010dz}) is sufficient to describe the glycerol-protein system. It does not necessarily mean that the same situation repeats itself for a hydrated protein, or applies equally well to the M{\"o}ssbauer experiment with a much longer resolution time of $t_r\simeq 140$ ns \cite{Parak:03}.  Some experimental data indeed claim the existence of independent relaxation processes of the protein hydration shells with significantly slower relaxation times \cite{Pal:04,Bhattacharyya:2008fk}. The resolution of this claim, however, depends on the water mode probed by the observations. There is a relatively insignificant slowing down of water's single-molecule rotational dynamics in hydration shells \cite{Laage:11}. An attempt to find a separate dynamic process in density fluctuations (translations) probed by depolarized light scattering resulted in the realization that cross protein-water correlations, instead of a separate dynamic process, can explain the data \cite{DMjcp2:14}. However, the collective variable of the shell dipole moment can be characterized as a separate dynamic process, which is  both significantly slower and is spatially extended into the bulk \cite{DMjcp1:12}. From a general perspective, a strong perturbation of the forces existing in the bulk is required for a new dynamic process to appear. If a significant alteration of the hydrogen-bond network is achieved in the solvation layer, one can expect a separate dynamic process to show up. The extent of such network perturbation is where the distinction between glycerol and water might be found.

\acknowledgments 
This research was supported by the National Science Foundation (CHE-1464810) and through XSEDE (TG-MCB080116N). We are grateful to Ranko Richert for his help with dielectric data for glycerol and for helpful suggestions on the manuscript. Discussing glass science with Austen Angell has inspired many of the questions raised in this paper.

\appendix

\section{Simulation protocol}
\label{secA1}
Molecular dynamics (MD) simulations were performed for twelve different temperatures (147, 168, 179, 195, 214, 239, 255, 275, 287, 302, 312, 334 K) in a cubic box consisting of 1000 glycerol molecules using the OPLS-AA (Optimized Potentials for Liquid Simulation - All Atoms) force field \cite{doi:10.1021/ct200731v} within the Gromacs \cite{Hess:2008db} simulation package. The initial box with 1000 glycerol molecules was downloaded from virtual chemistry website \cite{10.1007/b10103,doi:10.1021/ct200731v}. After the initial NPT and NVT equilibration runs, the raw box was used to produced 50 ns equilibration trajectories in the NVE ensemble with no constraints. The output was then used to generate the NVE trajectories. 

Each system was initialized with a 300 ps NVT run using a Nose-Hoover thermostat with H-bonds constrained followed by a 300 ps run with no constraints. Then a 1-3 ns NVE run was generated to check for stability before doing the 50 ns production run for each temperature. The time step for all production runs was 0.5 fs, with all atoms (including hydrogens) free to move according to the OPLS-AA force field parameters. The group cutoff-scheme was used with an update time of 5 ns and a cutoff distance of 1.1 nm for the shifted Lennard-Jones and electrostatic interactions with a group list distance of 13 \AA\ renewing every 10 simulation steps. Long-ranged electrostatic interactions were calculated with the particle mesh Ewald method. Additional trajectories (tens of ns) were generated for a separate raw box under the NVT ensemble as a means to compare the results between NVE and NVT ensembles. These NVT simulations with the H-bond constraints were equilibrated with an initial 3 ns run and an additional 3 ns with the constraints removed, assuring minimization and equilibration before production runs for each temperature were executed. The 50 ns NVT simulations were carried out for temperatures 230, 240, 250, 260, 270, 280 K. In this case, the Verlet cutoff-scheme was implemented and a Nose-Hoover thermostat was used. All simulations were carried out using periodic boundary conditions.

\bibliography{chem_abbr,dielectric,dm,statmech,glass,protein,liquids,solvation,dynamics,simulations,surface,water,nano,lcold,ferro,./SupplementGNotes1}

\end{document}